\documentstyle[12pt]{article}
\begin{document}
\baselineskip = 0.24 in		
\voffset = -30 mm		
\hoffset = -11 mm		
\raggedbottom
\renewcommand{\floatpagefraction}{0.95}
\renewcommand{\topfraction}{0.75}
\renewcommand{\textfraction}{0.25}
\normalsize
\rightline{CMU-HEP91-24-R92}
\vskip 0.3 in
\centerline{ISOSPIN SPLITTING IN THE BARYON OCTET AND DECUPLET}
\vskip 0.3 in
\centerline{R. E. Cutkosky}
\centerline{{Physics Department, Carnegie-Mellon University}}
\centerline{{Pittsburgh, PA 15213, USA}}
\vskip 0.4 in

\normalsize

\centerline{Abstract}
\vskip 0.20 in
\baselineskip = 0.2 in

Baryon mass splittings are analyzed in terms of a simple model with
general pairwise interactions.  At present, the $\Delta$ masses are
poorly known from experiments.   Improvement of these data would
provide an opportunity to make a significant test of our understanding
of electromagnetic and quark-mass contributions to hadronic masses.
The problem of determining resonance masses from scattering and
production data is discussed.

\baselineskip = 0.24 in		

\normalsize
\vskip 0.20 in

The isospin splittings of the masses and coupling constants of baryons
and mesons arise from the mass differences of the up and down quarks and
the electromagnetic interactions between them. These splittings provide
a good way to test our knowledge of the internal wave functions of these
particles.  At present there is no evidence for isospin splitting of
coupling constants, at least at the 3\% level, although further work
might show an effect\cite{gh}.  However, mass splittings in many
isospin multiplets are known, and there has been extensive theoretical
discussion.

I shall outline here a simple way to describe baryon splittings that
embodies the main features of more detailed calculations made by
others\cite{ni,sc}.  The explicit models often contain several
adjustable parameters that may not have a transparent meaning. One
advantage of rewriting the mass perturbation effects is that it provides
a clearer picture of how our theoretical understanding is affected by
the experimental situation.  A surprising aspect of this is that the
$\Delta$ masses are the ones that are in most need of further study.

To introduce the approach that is used here, and also to obtain some
parameters, I first consider the SU(6) splittings induced by hyperfine
interactions and by the difference $m_s-m_n$ of the strange quark mass
and the average of the natural quark masses.  The model is based on the
one introduced by De Rujula, Georgi, and Glashow\cite{DGG}.  I assume
the energy is the sum of one-body and two-body quark effects, in which
the third quark is an inert spectator. The one-body effects (simple mass
and kinetic energy terms) are included in the two-body effects, with
half of each single-quark term ascribed to each of two pairs.  There are
five different two-body terms distinguished by their quark content,
triplet terms $T_{nn}$, $T_{ns}$, and $T_{ss}$, along with singlet terms
$S_{nn}$ and $S_{ns}$.  In the simple $S$-wave model, decuplet states
contain 3 triplet pairs, while octet states contain $\frac{3}{2}$
triplet pairs and $\frac{3}{2}$ singlet pairs.  Some generalizations
of this picture are discussed later.

In fitting to the experimental quantities, I use a formulation in
which a ``model error'' $t$ is added in quadrature to the experimental
errors\cite{FC}.  This allows a consistent way to judge the general
goodness of fit in a situation in which some data are known much more
precisely than others.  If the model is not good enough to fit all data
within their errors, the model error is defined to be the value of $t$
required to give a value for $\chi^2$ equal to the number of degrees of
freedom.

The best fit to the central masses of isospin multiplets, which has a
model error of 8.9 MeV, is given in table \ref{multiplet}.  (These
central masses are obtained by fitting to isospin splittings.)
\begin{table}
\caption{Multiplet Masses}
\label{multiplet}
\vskip 0.1 in
\begin{tabular}{@{\hspace{0.03 in}}l
@{\hspace{0.27 in}}r@{\hspace{0.27 in}}r@{\hspace{0.27 in}}r
@{\hspace{0.27 in}}r@{\hspace{0.27 in}}r@{\hspace{0.27 in}}r
@{\hspace{0.27 in}}r@{\hspace{0.27 in}}c@{\hspace{0.27 in}}c}
        & $T_{nn}$ & $T_{ns}$ & $T_{ss}$ & $S_{nn}$ &  $S_{ns}$
        &total     &data   &$\pm$ &$\chi^2$\\[0.1in]
$N$     &618.4     & 0.0   &   0.0   & 318.7   & 0.0
        &937.0    &938.9   &0.00  &0.05\\
$\Lambda$ & 0.0   &730.2   &  0.0    & 212.5   & 175.7
        &1118.4   &1115.6   &0.50  &0.10\\
$\Sigma$  &412.2  &243.4   &   0.0   &  0.0    &527.2
        &1182.8   &1193.2   &0.04  &1.35\\
$\Xi$     & 0.0   &243.4   &556.8    &  0.0    &527.2
          &1327.4   &1318.0   &0.16  &1.12\\[0.1in]
$\Delta$    &1236.7 &0.0   &0.0   &0.0   &0.0
           &1236.7   &1232.8   &0.26  &0.20\\
$\Sigma^*$  &412.2 &973.6  &0.0   &0.0   &0.0
           &1385.8   &1384.7   &0.30  &0.02\\
$\Xi^*$     &0.0   &973.6 &556.8  &0.0   &0.0
           &1530.4   &1533.4   &0.36  &0.11\\
$\Omega$    &0.0   &0.0   &1670.4 &0.0   &0.0
           &1670.4   &1672.5   &0.30  &0.06
\end{tabular}
\end{table}
It is seen that this simple pair model can fit the splittings to
within 5\%. The $\Sigma$ and $\Xi$ masses are the most discrepant. The
pair term energies that give this fit, shown in table \ref{pair}, will
be used later to estimate parameters for a more explicit model.
\begin{table}
\caption{Pair Energies}
\label{pair}
\vskip 0.1 in
\begin{tabular}{@{\hspace{2.15 in}}cccc}
    &  $nn$  & $ns$  & $ss$  \\
$T$ & 412.25 &486.80 &556.79 \\
$S$ & 212.46 &351.47    &
\end{tabular}
\end{table}
In the context of this model, it is possible to interpret the triplet
contributions as containing contributions of tensor forces, to the
extent that these do not lead to non-spectator effects.  It is
possible to add plausible non-spectator terms that reduce the model
error, but this is not useful here.

To fit the masses of the individual isospin components, besides the
central mass values, there are four distinct isospin-splitting pair
terms to be used.  These are $T^1$, $T^2$, $T^1_s$, and $S^1_s$, where
the superscript $I$ denotes the isospin tensorial rank and a subscript
$s$ indicates that the pair contains one strange quark.  Note that the
Coleman-Glashow relation\cite{CG} among octet masses is automatically
satisfied by parametrization with the three independent $I=1$ pair
terms, independently of any specific model of the origin of the
isospin-splitting terms.  In addition, however, the pair model implies
that relations exist among the octet and decuplet mass splittings. Use
of all available data from the PDG compilation\cite{PDG} except the
$\Delta^+$ mass gives a fit with a model error of 0.14 MeV, which is
also about 5\% of typical splittings. This fit is the main result
reported here, and provides a starting point for further discussion.
The numerical values, identified here as fit $A$, are shown in table
\ref{ipair} and in the figures.
\begin{table}
\caption{Contributions to particle masses in fit $A$}
\label{ipair}
\vskip 0.05 in
\begin{tabular}{@{\hspace{0.05 in}}l
@{\hspace{0.21 in}}r@{\hspace{0.21 in}}r@{\hspace{0.21 in}}r
@{\hspace{0.21 in}}r@{\hspace{0.21 in}}r@{\hspace{0.21 in}}r
@{\hspace{0.21 in}}r@{\hspace{0.21 in}}c@{\hspace{0.21 in}}c}
 &    $ M_0$  &   $T^1$  &  $T^2$ &  $T_s^1$  &   $S_s^1$ &total
      &  data  &  $\pm$ &  $\chi^2$ \\[0.1in]
$N^+$ &   938.92  &$-$0.74   &0.00   &0.00   &0.00   &938.18
      &938.27  &0.00  &0.39 \\
$N^0$ &   938.92   &0.74   &0.00   &0.00   &0.00   &939.66
      &939.57  &0.00  &0.39 \\[0.1in]
$\Sigma^+$ &1193.15  &$-$0.74   &0.24  &$-$0.39  &$-$2.94  &1189.32
           &1189.37  &0.07  &0.09 \\
$\Sigma^0$ &1193.15   &0.00  &$-$0.47   &0.00   &0.00  &1192.68
           &1192.55  &0.10  &0.50 \\
$\Sigma^-$ &1193.15   &0.74   &0.24   &0.39   &2.94  &1197.45
           &1197.50  &0.05  &0.11 \\[0.1in]
$\Xi^0$   &1318.02   &0.00   &0.00  &$-$0.39  &$-$2.94  &1314.69
          &1314.80  &0.80  &0.02 \\
$\Xi^-$   &1318.02   &0.00   &0.00   &0.39   &2.94  &1321.35
          &1321.34  &0.14  &0.00 \\[0.2in]

$\Delta^{++}$ &1232.76  &$-$2.21   &0.71   &0.00   &0.00  &1231.25
              &1231.00  &0.30  &0.58 \\
$\Delta^+$    &1232.76  &$-$0.74  &$-$0.71   &0.00   &0.00  &1231.31
              &1234.90  &1.40  &* \\
$\Delta^0$    &1232.76   &0.74  &$-$0.71   &0.00   &0.00  &1232.78
              &1233.40  &0.50  &1.41 \\
$\Delta^-$    &1232.76   &2.21   &0.71   &0.00   &0.00  &1235.68
              &?$\quad\,$ &$\infty$  &* \\[0.1in]

$\Sigma^{*+}$  &1384.72  &$-$0.74   &0.24  &$-$1.54   &0.00  &1382.68
               &1382.80  &0.40  &0.08 \\
$\Sigma^{*0}$  &1384.72   &0.00  &$-$0.47   &0.00   &0.00  &1384.25
               &1383.70  &1.00  &0.29 \\
$\Sigma^{*-}$  &1384.72   &0.74   &0.24   &1.54   &0.00  &1387.24
               &1387.20  &0.50  &0.01 \\[0.1in]

$\Xi^{*0}$     &1533.38   &0.00   &0.00  &$-$1.54   &0.00  &1531.84
               &1531.78  &0.34  &0.02 \\
$\Xi^{*-}$     &1533.38   &0.00   &0.00   &1.54   &0.00  &1534.93
               &1535.20  &0.80  &0.11
\end{tabular}
\vskip 0.05 in

\hspace{0.1 in}* = omitted from fit

\end{table}

The results in table \ref{ipair} show that the $\Delta$ masses are the
hardest to fit.  The data used here are an average of the values
obtained by Koch and Pietarinen\cite{KP} and Abaev\cite{va} (which are
very similar). The VPI group{\cite{raa} has recently obtained the
preliminary value $M(\Delta^0) - M(\Delta^{++}) = 1.1 \pm 0.1$ MeV,
which is more compatible with the model.  The unused discrepant
$\Delta^+$ datum was obtained from photoproduction\cite{iim}.
Crawford\cite{rlc} has reported a value (1231.6 MeV) that is more
consistent with the prediction, but did not estimate the error.  If all
$\Delta$ masses are omitted from the input, the resulting fit ($B$)
needs no model error.  The predicted masses are shown in table
\ref{Delta}. They deviate slightly more from the data than do the
values from the combined fit $A$.
\begin{table}
\caption{Predicted $\Delta$ masses from fit $B$}
\label{Delta}
\vskip 0.1 in
\begin{tabular}{@{\hspace{0.8 in}}lcrrcccc}
              &$M_0$   &$T^1$   &$T^2$     &$B$   &$A$  &data  &$\pm$ \\
$\Delta^{++}$  &1232.48  &$-$1.94   &0.89   &1231.43  &1231.25  &1231.0
&0.3 \\
$\Delta^+$     &1232.48  &$-$0.65  &$-$0.89   &1230.95  &1231.31  &1234.9
&1.4 \\
$\Delta^0$     &1232.48   &0.65  &$-$0.89   &1232.24  &1232.78  &1233.4
&0.5  \\
$\Delta^-$     &1232.48   &1.94   &0.89   &1235.31  &1235.68  &    ?
&$\infty$
\end{tabular}
\end{table}

\setlength{\unitlength}{1 mm}
\begin{figure}
\begin{picture}(160,120)(-15,-10)
\thicklines
\multiput(0,10)(0,10){9}{\line(1,0){3}}
\put(0,0){\framebox(130,100)}
\put(0,50){\line(1,0){130}}
\put(-5,49){0}
\put(-5,69){2}
\put(-5,89){4}
\put(-7,29){$-2$}
\put(-7,9){$-4$}
\multiput(30,0)(10,0){2}{\line(0,1){2}}
\put(28,-7){$N^+$}
\put(25,43.5){\line(1,0){10}}
\put(30,42.6){\circle{2.8}}
\put(38,-7){$N^0$}
\put(35,56.5){\line(1,0){10}}
\put(40,57.4){\circle{2.8}}
\multiput(60,0)(10,0){3}{\line(0,1){2}}
\put(58,-7){$\Sigma^+$}
\put(55,12.2){\line(1,0){10}}
\put(60,11.5){\line(0,1){1.4}}
\put(60,11.7){\circle{2.8}}
\put(68,-7){$\Sigma^0$}
\put(65,44.0){\line(1,0){10}}
\put(70,43.0){\line(0,1){2.0}}
\put(70,45.3){\circle{2.8}}
\put(78,-7){$\Sigma^-$}
\put(75,93.5){\line(1,0){10}}
\put(80,93.0){\line(0,1){1.0}}
\put(80,93.0){\circle{2.8}}
\multiput(100,0)(10,0){2}{\line(0,1){2}}
\put(98,-7){$\Xi^0$}
\put(95,17.8){\line(1,0){10}}
\put(100,9.8){\line(0,1){16.0}}
\put(100,16.7){\circle{2.8}}
\put(108,-7){$\Xi^-$}
\put(105,83.2){\line(1,0){10}}
\put(110,81.8){\line(0,1){2.8}}
\put(110,83.3){\circle{2.8}}
\end{picture}
\caption{Octet isospin splittings, in MeV.  The crosses give the
experimental data and errors, and the circles show the fitted values.
The radius is given by the model error.}
\label{octet}
\end{figure}
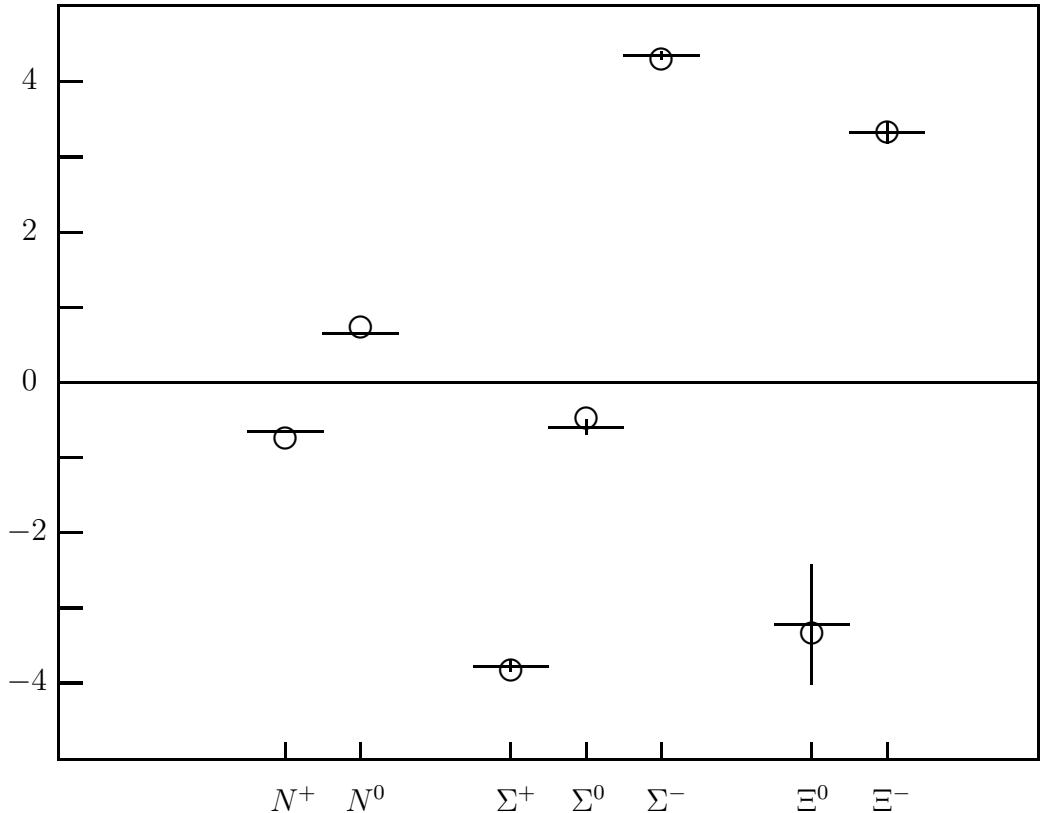

Pedroni, {\it et al.}, have measured the total cross sections for
scattering of $\pi^{\pm}$ mesons from deuterium\cite{Ped}. In the
impulse approximation, this can determine a value for the mass
combination $D = M(\Delta^-) - M(\Delta^{++})
+\hbox{$\frac{1}{3}$}(M(\Delta^0)-M(\Delta^+) )$. Their result, after
numerous and sizeable theoretical corrections, was $D= 4.6 \pm 0.2$
MeV, which corresponds to $-T^1(\Delta)=1.38 \pm 0.06$ MeV. The
uncertainty here represents only the statistical errors.  This value,
obtained by experimentation involving only $\Delta$ states, is
intermediate between the values obtained from the more global fits $A$
and $B$.

Although the ease of fitting with the model was represented by a
``model error'', it was actually the experimental $\Delta$ masses that
were hard to accommodate. These experimentally-derived values may also
be subject to unrecognized model-dependent systematic errors --- this
is a separate question. The model error was used here as a device to
allow the $N$ and $\Sigma$ masses to relax somewhat from their
precisely-known values in a global fit. Even so, it was the $\Delta$
masses that contributed the most to $\chi^2$, and when the $\Delta$
masses were ignored, the other mass values did not need to relax.

In a simple $S$-wave picture of the states, using first-order
perturbation theory, the four pair energies can be expressed as linear
combinations of four distinct contributions.  The first of these is the
single-particle mass term $\delta=\kappa(m_d-m_u)$, where $\kappa$ is a
reduction factor arising from the momentum of the quarks. The
``constituent" masses are interpreted here as ``magnetic moment" masses.
A simple standard fit to magnetic moments gives $m_n \sim 344$ MeV and
$m_s \sim 533$ MeV, and $x = m_n/m_s = 0.65$.  A second effect of the
mass differences arises from the color-hyperfine interaction,
parametrized by a coefficient $d$.  There is also a Coulomb term $C$,
proportional to $\langle 1/r \rangle$, and a magnetic interaction term
with a coefficient $b$.  Both $b$ and $d$ are proportional to
$\psi(0)^2$, and inversely proportional to the product of masses. Taking
into account charge and spin factors, the pair energies can be expressed
as
\begin{eqnarray}
\label{epair}
T^1& = &  \hbox{$\frac{1}{3}$} C -\hbox{$\frac{1}{3}$}
b -\delta + 2d \, , \nonumber \\
T^2& = & C - b \, , \nonumber \\
T^1_s & = & -\hbox{$\frac{1}{3}$}
 C +\hbox{$\frac{1}{3}$} xb -\hbox{$\frac{1}{2}$} \delta +xd \, , \nonumber \\
S^1_s & = & -\hbox{$\frac{1}{3}$} C - xb -\hbox{$\frac{1}{2}$}
 \delta -3xd \, .
\end{eqnarray}

Stevenson, {\it et al.}, (SMG)\cite{SMG} have pointed out that
electromagnetic box and penguin graphs can contribute additional
effects.  However, they consider these only within the general class
of pair terms, and these graphs do not alter the fact that there can
be only four such independent terms.  Rather, they give additional
contributions to the right hand side of Eq.\ (\ref{epair}), and it
would not be possible to determine them independently from the data.
These effects call attention to the fact that the effective mass
differences of quarks subjected to confinement have themselves been
influenced by electromagnetic contributions.  In other words, a clean
separation of electromagnetic from quark-mass effects is not really
possible. In particular, this affects the interpretation of the quark
mass parameters $\delta$ and $d$.  The meaning ascribed to the
quantities $\langle 1/r \rangle$ and $\psi(0)^2$ could also be
changed. The additional contributions introduced by SMG, as well as
the four simple effects included in Eq.\ (\ref{epair}), may also
depend on the environment provided by the third quark and thus also
contribute to non-spectator effects.

In the simplified model, it is not possible to determine the four
parameters from Eq.\ (\ref{epair}) because the coefficient matrix is
singular. In this model, the pair energies should satisfy the
constraint
\begin{equation}
\label{constraint}
2xT^1-2xT^2-(3x+1)T^1_s+(1-x)S^1_s =0 \, .
\end{equation}
Using $x=\frac{2}{3}$ and the covariance matrix from fit $A$ gives for
the LHS of Eq.\ (\ref{constraint}) the value $-0.54 \pm 1.07$ MeV. The
constraint is therefore acceptable, and there is no need for
additional terms. To eliminate $d$ as a free parameter, let
$d=\tau\delta$ and use the same model to determine $\tau$ from the
pair energies given in table \ref{pair}. The mass-dependent hyperfine
effect can be isolated in the quantity
\begin{eqnarray}
\label{D}
d_s &= &\hbox{$\frac{1}{4}$} \left( T_{nn} - S_{nn} -T_{ns} + S_{ns}
\right) \nonumber \\
     &= & 16.1 \pm 2.1\, \hbox{MeV} \, .
\end{eqnarray}
The pair energies also provide estimated values $x = 0.68$ and $\delta_s
= \kappa_s (m_s - m_n) = 181$ MeV, which are compatible with the values
from magnetic moments. The estimate $\kappa_s/\kappa \sim 1.5$ then
gives
\begin{equation}
\label{tau}
\tau = \frac{\kappa_s d_s}{x\kappa\delta_s} = 0.2 \, .
\end{equation}
Fitting this constrained model to the pair energies from fit $A$ gives
an acceptable $\chi^2$ and the results shown in table \ref{model}.
\begin{table}
\caption{Model contributions to pair energies}
\label{model}
\vskip 0.05 in
\begin{tabular}{@{\hspace{1.45 in}}lrrrrrr}
        &$C$     &$b$  &$\delta$   &$d$   &total   &input \\
$T^1$   &0.66   &--0.22  &--3.13   &1.25  &--1.43    &--1.48  \\
$T^2$   &1.97   &--0.64   &0.0     &0.0     &1.33     &1.42  \\
$T^1_s$  &--0.66   &0.14   &--1.56   &0.42   &--1.66    &--1.54  \\
$S^1_s$  &--0.66   &--0.43  &--1.56   &--1.25  &--3.90    &--3.92
\end{tabular}
\end{table}
The numerical values of the parameters are similar to those determined
by Isgur\cite{ni}. Note that there is considerable cancelation,
especially in $T^1$, and except in $S^1_s$.  This helps to explain why
isospin spittings for hyperons are much larger than for the $N$ and
$\Delta$.  It also suggests that the latter may be more sensitive to
non-spectator effects, because the pair effects tend to cancel, and
the corrections might not.

\begin{figure}
\begin{picture}(160,100)(-10,100)
\thicklines
\put(0,110){\framebox(140,80)}
\multiput(0,120)(0,10){7}{\line(1,0){3}}
\put(0,150){\line(1,0){140}}
\put(-5,149){0}
\put(-5,169){2}
\put(-7,129){$-2$}
\multiput(20,110)(10,0){4}{\line(0,1){2}}
\put(17,103){$\Delta^{++}$}
\put(15,132.4){\line(1,0){10}}
\put(20,129.4){\line(0,1){6.0}}
\put(20,134.9){\circle{2.8}}
\put(20,139.5){\circle*{1.5}}
\put(28,103){$\Delta^+$}
\multiput(25,171.4)(4.0,0){3}{\line(1,0){2}}
\multiput(30,157.4)(0.,4.3){7}{\line(0,1){2.1}}
\put(30,135.5){\circle{2.8}}
\put(30,134.7){\circle*{1.5}}
\put(38,103){$\Delta^0$}
\put(35,156.4){\line(1,0){10}}
\put(40,151.4){\line(0,1){10.0}}
\put(40,150.2){\circle{2.8}}
\put(40,147.6){\circle*{1.5}}
\put(48,103){$\Delta^-$}
\put(50,179.2){\circle{2.8}}
\put(50,178.3){\circle*{1.5}}
\multiput(70,110)(10,0){3}{\line(0,1){2}}
\put(67,103){$\Sigma^{*+}$}
\put(65,130.8){\line(1,0){10}}
\put(70,126.8){\line(0,1){8.0}}
\put(70,129.6){\circle{2.8}}
\put(77,103){$\Sigma^{*0}$}
\put(75,139.8){\line(1,0){10}}
\put(80,129.8){\line(0,1){20.0}}
\put(80,145.3){\circle{2.8}}
\put(87,103){$\Sigma^{*-}$}
\put(85,174.8){\line(1,0){10}}
\put(90,169.8){\line(0,1){10.0}}
\put(90,175.2){\circle{2.8}}
\multiput(110,110)(10,0){2}{\line(0,1){2}}
\put(107,103){$\Xi^{*0}$}
\put(105,134.0){\line(1,0){10}}
\put(110,130.6){\line(0,1){6.8}}
\put(110,134.6){\circle{2.8}}
\put(117,103){$\Xi^{*-}$}
\put(115,168.2){\line(1,0){10}}
\put(120,160.2){\line(0,1){16.0}}
\put(120,165.5){\circle{2.8}}
\end{picture}
\caption{Decuplet isospin splittings, in MeV.  The small filled circles
give the predicted $\Delta$ masses, when the $\Delta$ data are omitted.}
\label{Decuplet}
\end{figure}
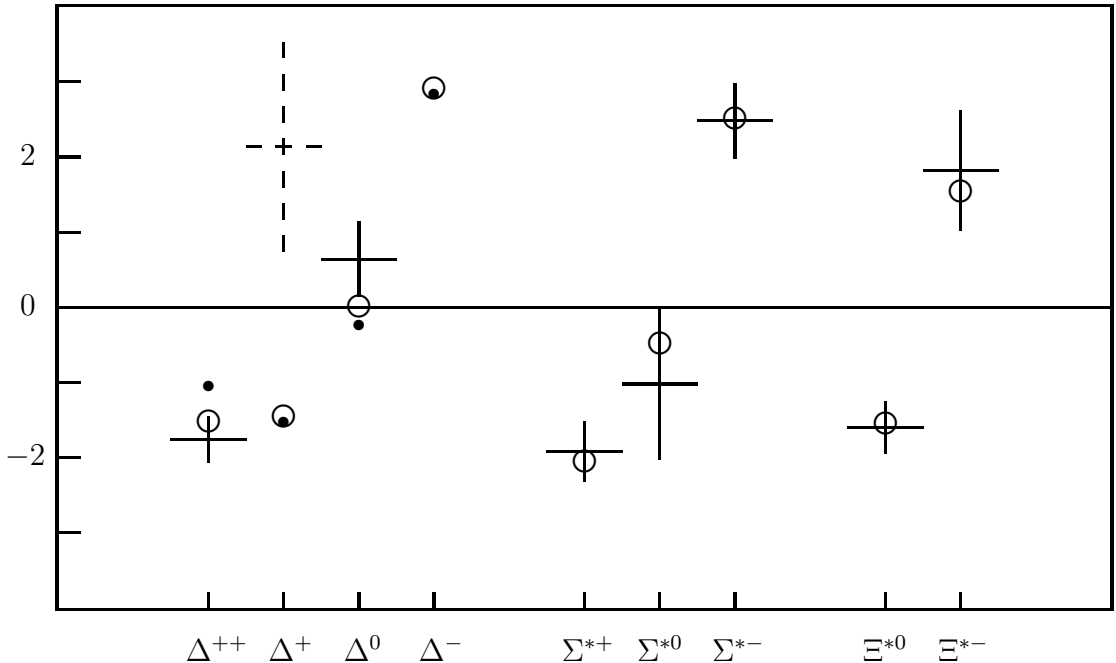

A comparison of the fitted values of $T^1$ and $T^2$ with the numbers
obtained from recent explicit baryon models is shown in table
\ref{triplet}.
\begin{table}
\caption{Comparison of triplet pair energies}
\label{triplet}
\vskip 0.1 in
\begin{tabular}{@{\hspace{2.0 in}}rcccc}
        & A      & B   &I &C  \\
$-T^1$ & 1.48 &    1.29 &   2.3    &  1.9 \\[0.05 in]
$T^2$ &   1.42 &   1.78 &   1.6    &  1.7
\end{tabular}

\vskip 0.05 in
\hspace{2.0 in}I = Isgur\cite{ni}, C = Capstick\cite{sc}

\end{table}
The Isgur\cite{ni} and Capstick\cite{sc} values for
$T^1$ and $T^2$ listed in this table were obtained from their
predictions for $\Delta$ masses.  They both adjusted some parameters
to fit the $n$ - $p$ mass difference $-T^1(N)=1.3$ MeV. The difference
between their $T^1$ values for the $N$ and $\Delta$ is a measure of
the non-spectator effects in their models.

The results in table \ref{triplet} show that the $\Delta$ masses
provide a sensitive test of models.  These masses also provide the best
opportunity for improvement in the experimental data.  At the same
time, it should be possible to study isospin breaking effects in their
partial widths.

A different view, that improved $\Sigma^*$ mass values would provide
the most improvement in our understanding, was expressed by
SMG\cite{SMG}.  The $\Sigma^*$ and $\Xi^*$ masses certainly have large
uncertainties, as can be seen from Fig.\ \ref{Decuplet}.  It is clear
that reduction of these uncertainties would contribute greatly to our
understanding of the structure of decuplet states. However, the
apparent sensitivity of the undetermined SMG parameters to the input
$\Sigma^*$ data also depends in part on the method of fitting they
employed.  Moreover, it has been shown here that the reported $\Delta$
masses already show an inconsistency with pair-interaction models that
do incorporate the terms suggested by SMG. There are also two $\Delta$
charge states for which reliable masses are not yet available.
Examination of the $\Delta$ mass determinations suggests that, in
addition to the statistical errors, these may at present be subject to
model-dependent systematic errors amounting to perhaps
\hbox{$\frac{1}{2}$} MeV.  This should be resolved by new, independent
analyses.  Similar effects would also be present in the $\Sigma^*$ and
$\Xi^*$ masses, but in these cases the existing statistical errors
dominate.

The explicit calculations\cite{ni,sc} suggest that decuplet masses may
be sensitive to model parameters that influence non-spectator
contributions. Independently, it might be expected that pair models
would fail to describe splittings in baryons that contain heavy
quarks. The structure of states that contain charm or bottom quarks
might differ in important respects from baryons with three lighter
quarks.  Further experimentation with heavy quark systems could give
useful information about this important point.

Improved experimentation and analysis for the reactions $\gamma p
\rightarrow \pi^0 p \, \,(\pi^+ n)$ could lead to significantly better
mass values for the $\Delta^+$.  In addition, further experimentation
with deuteron targets could provide valuable information about all the
$\Delta$ states, but would also require difficult and careful
analysis. Quasi-free scattering of $\pi^-$ from the neutron could give
information about the $\Delta^-$. A set of experiments comparing
quasi-free $\pi^{\pm}$ scattering from the proton and the neutron in
the deuteron would provide a set of interlocking comparisons in which
some of the systematic errors in the determination of the free-nucleon
cross sections might cancel out. Elastic differential $\pi^{\pm}$
scattering from deuterons would be more difficult to analyse, but
would give another way to compare amplitudes, especially for the
$\Delta^{++}$ and the $\Delta^{-}$. Similar analyses could be applied
to photoproduction from the deuteron, and give a direct comparison of
$\Delta^{+}$ with $\Delta^{0}$.

Comparison of the masses and couplings of states with different flavors
requires use of a common set of conventions and definitions.  This is
especially important when data from different kinds of experiments are
used.  Consider the $S$-matrix for an isolated narrow multichannel
resonance, which may be written as $S(E)=\tilde{S}_BS_R(E)S_B$, where $S_R$
is a simple Breit-Wigner resonance factor (with energy-dependent partial
widths) and $S_B$ is a slowly varying background factor.  In analysis of
production experiments, the initial factor $S_B$ would be replaced by
other factors depending on energy.  In general, the resonance energy
appearing in the factor $S_R$ provides a satisfactory initial estimate
for the mass of the excited resonant state.  However, removal of the
background amplitude may introduce some model dependence.  One way to
proceed would be to look for the resonance pole, and then use the
explicit energy dependence in $S_R$ to go back to the real axis.

At first sight, the problem of accounting for electromagnetic
corrections to scattering amplitudes is relatively straightforward, but
it involves several distinct aspects. The first aspect, that of
correctly calculating the contribution of the initial and final state
electromagnetic interactions to the $\pi p$ scattering amplitudes, can
be treated by the method of Tromborg, {\it et al.}\cite{bt}.  However,
the best correction to be applied to determine the effective energy at
which hypothetical chargeless particles would interact has not been
firmly established. Furthermore, the models considered here involve
idealized states with three quarks.  The coupling to baryon-meson states
introduces certain mass shifts.  What we are concerned with here is the
part of the difference in these shifts that originates in the
electromagnetic interactions and the mass differences in the coupled
channels, and it may be fruitful to focus attention on this restricted
problem.  These questions also arise in discussions of the rest of the
baryon decuplet, and should be given a uniform treatment.

\vskip 0.10 in
\noindent {\bf Acknowledgements:} I wish to thank S. Capstick for
information about isospin breaking models, G. H\"ohler for
communications about the problem of extracting resonance parameters
from data, and V. Abaev and R. Arndt for discussions of their
analyses. I thank a referee for calling my attention to the relevance
of the paper by Stevenson, {\it et al.} This work was supported by the
U. S. Dept.\ of Energy under contract No.\ DE-AC02-76ER-03066.

\end{document}